\documentclass[aps,preprint,amsmath,tightenlines]{revtex4}

\usepackage{graphicx}
\usepackage{bm}

\begin{document}

\title{Implications of measured properties\\ of the mixing matrix
on mass matrices}

\author{S.~Chaturvedi}

\email[]{scsp@uohyd.ernet.in}

\affiliation{
School of Physics.\\
University of Hyderabad\\
Hyderabad 500 046, India.
}

\author{V.~Gupta}

\email[]{virendra@mda.cinvestav.mx}

\author{G.~S\'anchez-Col\'on}

\email[]{gsanchez@mda.cinvestav.mx}

\affiliation{
Departamento de F\'{\i}sica Aplicada.\\
Centro de Investigaci\'on y de Estudios Avanzados del IPN.\\
Unidad Merida.\\
A.P. 73, Cordemex.\\
M\'erida, Yucat\'an, 97310. MEXICO.
}

\date{\today}

\begin{abstract}

It is shown how the two experimentally measurable  properties of
the mixing matrix $V$, the asymmetry $\Delta(V)=|V_{12}|^2-
|V_{21}|^2$ of $V$ with respect to the main diagonal and the
Jarlskog invariant $J(V)={\rm Im}(V_{11}V_{12}^*V_{21}^*V_{22})$,
can be exploited to obtain constraints on possible structures of
mass matrices in the quark sector. Specific mass matrices are
examined in detail as an illustration.

\end{abstract}


\maketitle

\section{\label{introduction}Introduction}

Flavor mixing in the quarks, in the Standard Model, arises from
the unitary mixing matrices which diagonalize the
corresponding mass matrices. In the quark sector~\cite{2}, in
the physical basis, the CKM-mixing matrix is given by $V=
V^\dagger_{\rm u}V_{\rm d}$, where the unitary matrices $V_{\rm
u}$ and $V_{\rm d}$ diagonalize the up-quark and down-quark mass
matrices, respectively. One can also work in a basis in which the
up-quark (down-quark) mass matrix $M_{\rm u}$ ($M_{\rm
d}$) is diagonal. In these bases the mixing matrix in the quark
sector (like in the neutrino sector) will come from a single
mass matrix. Clearly, if we knew the mass matrices fully then
the corresponding mixing matrices are completely determined. In
practice, the mass matrices are guessed at, while experiment can
only determine the modulii of the matrix elements of the mixing
matrix.

Recently it was shown~\cite{1} that a general property
of the diagonalizing unitary matrix $U$ imply constraints on the
corresponding hermitian mass matrix $M$. In particular, it was
shown that the asymmetry $\Delta(U)$ w.r.t.\ the main diagonal
and the Jarlskog invariant $J(U)$~\cite{4}, which is a measure
of CP-violation, can be directly expressed in terms of the
eigenvalues $m_i$ and matrix elements $M_{ij}$ of the mass matrix
$M$. Since $U^\dagger M\,U=\hat{M}={\rm diag}[m_1,m_2,m_3]$ one
obtains

\begin{eqnarray}
\Delta(U)&\equiv & |U_{12}|^2-|U_{21}|^2=|U_{23}|^2-|U_{32}|^2=
|U_{31}|^2-|U_{13}|^2
\nonumber \\
&=&\frac{1}{D(m)}\{\sum_{k}\left(m_k~(M^2)_{kk}-m_{k}^{2}~M_{kk}\right)
\},
\label{7}
\end{eqnarray}

\noindent
where

\begin{equation}
D(m)\equiv
\begin{vmatrix}
1 & 1 & 1 \\
m_1 & m_2 & m_3 \\
 m_1^2 & m_2^2 & m_3^2
\end{vmatrix} = (m_2-m_1)(m_3-m_1)(m_3-m_2)
\label{8}
\end{equation}

\noindent
and

\begin{equation}
J(U)\equiv {\rm Im}(U_{11}U_{12}^* U_{21}^* U_{22})
=\frac{{\rm Im}(M_{12}M_{23}M_{13}^*)}{D(m)}.
\label{12}
\end{equation}

Also, in terms of $M$ and its eigenvalues,

\begin{equation}
|U_{k\alpha}|^2=(N_\alpha)_{kk},
\label{5}
\end{equation}

\noindent
where

\begin{equation}
N_\alpha=\frac{(m_\beta-M)(m_\gamma -M)}{(m_\beta-m_\alpha)
(m_\gamma -m_\alpha)},~~~\alpha\neq \beta \neq \gamma,
\label{4}
\end{equation}

\noindent
with $\alpha$, $\beta$, $\gamma$ taking values from 1 to 3.
Through this equation each $|U_{k\alpha}|$ can be calculated in
terms of the eigenvalues (assuming non-degenerate eigenvalues
which is true for the quarks) and matrix elements of $M$. Then,
$|U_{k\alpha}|$ so calculated will automatically satisfy the
unitarity relations $\sum_{k}|U_{k\alpha}|^2 = 1 =
\sum_{\alpha}|U_{\alpha k}|^2$. Thus, the calculated $\Delta(U)$
will be unique.

Eqns.~$(\ref{7})$ and $(\ref{12})$ provide a simple criterion
for selecting suitable mass matrices. In particular, the latter
is remarkable in that it shows that if $M_{12}M_{23}M_{13}^*$ is
real for a given $M$, then the Jarlskog invariant for the matrix
$U$ which diagonalizes it vanishes.

\section{\label{choice}Choice for the mass matrix $\bm{M}$}

We consider

\begin{equation}
M=
\begin{pmatrix}
0 & a & d \\
a^* & 0 & b \\
d^* & b^* & c
\end{pmatrix}
.
\label{13}
\end{equation}

\noindent
For $d=0$ this reduces to the so called Fritzsch type mass
matrix~\cite{7,8} and will give $J(U)=0$. We now
investigate its viability in both the up-quark and down-quark
diagonal bases.

From the characteristic equation, we have

\begin{eqnarray}
c&=&m_1+m_2+m_3,
\label{14}\\
-(|a|^2+|b|^2+|d|^2)&=&m_1m_2+m_1m_3+m_2m_3,
\label{15}\\
-c|a|^2+2{\rm Re}(abd^*)&=&m_1m_2m_3.
\label{16}
\end{eqnarray}

\noindent
For the quark sector we need the mass hierarchy $|m_1|<<|m_2|<<|m_3|$. This
coupled with Eqs.~$(\ref{15})$ and $(\ref{16})$ require
$m_1,m_3>0$ and $m_2<0$, assuming $c>0$, for both up and down
quarks. For simplicity we take $a$ and $b$ to be real and
positive and $d$ as pure imaginary. Eq.~$(\ref{16})$
then determines $a$. Eq.~$(\ref{12})$ gives $ab|d|$. This
together with Eq.~$(\ref{15})$ fixes $b$ and $|d|$.

\subsection{\label{casea}Down-quark diagonal basis}

In this case $M=M_{\rm u}$ is the up-quark mass matrix which is
diagonalized by $V_{\rm u}$. So the CKM-matrix $V=V^\dagger_{\rm
u}$ since $V_{\rm d}=I$. Note that $\Delta(V_{\rm u})=-\Delta(V)$
and $J(V_{\rm u})=-J(V)<0$, so we choose $d=-i|d|$ in this case.
For $J(V)$ and the quark masses we take the experimental values
given in~\cite{3}: $J(V)=(3.08\pm 0.18)\times 10^{-5}>0$,
$m_{\rm u}=(2.25\pm 0.75)\,{\rm Mev}$, $|m_{\rm c}|=(1.25\pm 0.09)\times
10^3\,{\rm Mev}$, and $m_{\rm t}=(174.2\pm 3.3)\times 10^3\,{\rm
Mev}$. Using these we obtain $D(m)=(-3.83\pm 0.31)\times
10^{13}\,({\rm Mev})^3$ and

\begin{eqnarray}
a & = & (53.5\pm 9.1)\,{\rm Mev},\quad
b=(14.67\pm 0.55)\times 10^3\,{\rm Mev},
\nonumber \\
c & = & (17.30\pm 0.33)\times 10^4\,{\rm Mev},\quad
|d|=(1.51\pm0.27)\times 10^3\,{\rm Mev}.
\label{18}
\end{eqnarray}

\noindent
Thus $M_{\rm u}$ is completely determined. We can calculate
$\Delta(V)$ and individual $|V_{k\alpha}|^2$ using
Eq.~$(\ref{7})$ and Eq.~$(\ref{5})$, respectively. The results
for are given in columns Case~A of Tables~\ref{table1}
and~\ref{table2}.

\subsection{\label{caseb}Up-quark diagonal basis}

In this case $M=M_{\rm d}$ is the down-quark mass matrix which is
diagonalized by $V_{\rm d}$. So the CKM-mixing matrix $V=V_{\rm
d}$ since $V_{\rm u}=I$. Here $\Delta(V_{\rm d})=\Delta(V)$
and $J(V_{\rm d})=J(V)>0$, so we choose $d=i|d|$ in this case. For
numerical analysis we take the values cited in~\cite{3} as
inputs: $J(V)=(3.08 \pm 0.18)\times 10^{-5}$, $m_{\rm d}=(5\pm
2)\,{\rm Mev}$, $|m_{\rm s}|=(95\pm 25)\,{\rm Mev}$, $m_{\rm b}=(4200\pm
70)\,{\rm Mev}$. These give $D(m)=(-1.80\pm 0.47)\times
10^9\,({\rm Mev})^3$ and

\begin{eqnarray}
a & = & (22.0\pm5.3)\,{\rm Mev},\quad
b=(615\pm 86)\,{\rm Mev},
\nonumber \\
c & = & (4110\pm 74)\,{\rm Mev},\quad
|d|=(4.10\pm 0.74)\,{\rm Mev}.
\label{19}
\end{eqnarray}

\noindent
Using these we can calculate $\Delta(V)$ and individual
$|V_{k\alpha}|^2$ as before. The results are given in columns
Case~B of Tables~\ref{table1} and~\ref{table2}, respectively.

\section{\label{deltaofv}Analisys of $\bm{\Delta(U)}$}

\begin{figure}
\includegraphics*{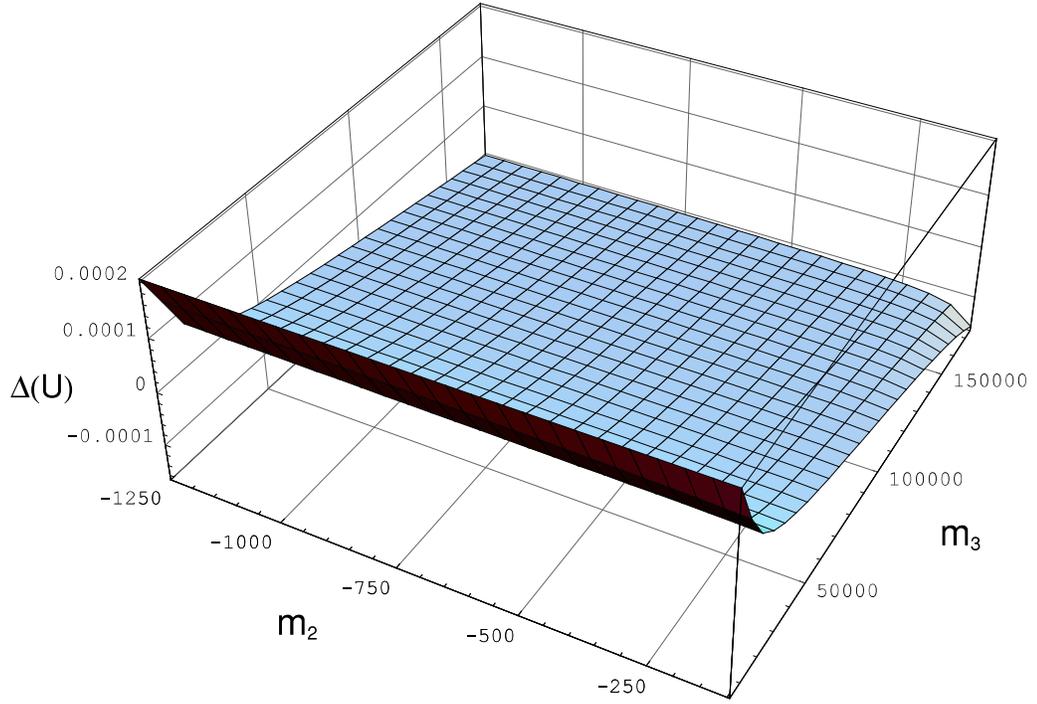}
\caption{\label{fig_1}Behaviour of $\Delta(U)$ as a function
of $m_2$ and $m_3$ for $m_1=1.5\,{\rm MeV}$. The range for $m_2$
is from $-m_s=-95\,{\rm MeV}$ to $-m_c=-1.25\,{\rm GeV}$ and for
$m_3$ from $m_b=4.20\,{\rm GeV}$ to $m_t=174.2\,{\rm GeV}$.
$\Delta(U)<0$ for the Case~A corner (down--quark diagonal basis)
where $|m_2|=m_c$ and $|m_3|=m_t$ and $\Delta(U)>0$ for the
Case~B corner (up--quark diagonal basis, $|m_2|=m_s$,
$|m_3|=m_b$).} \end{figure}

\begin{figure}
\includegraphics*{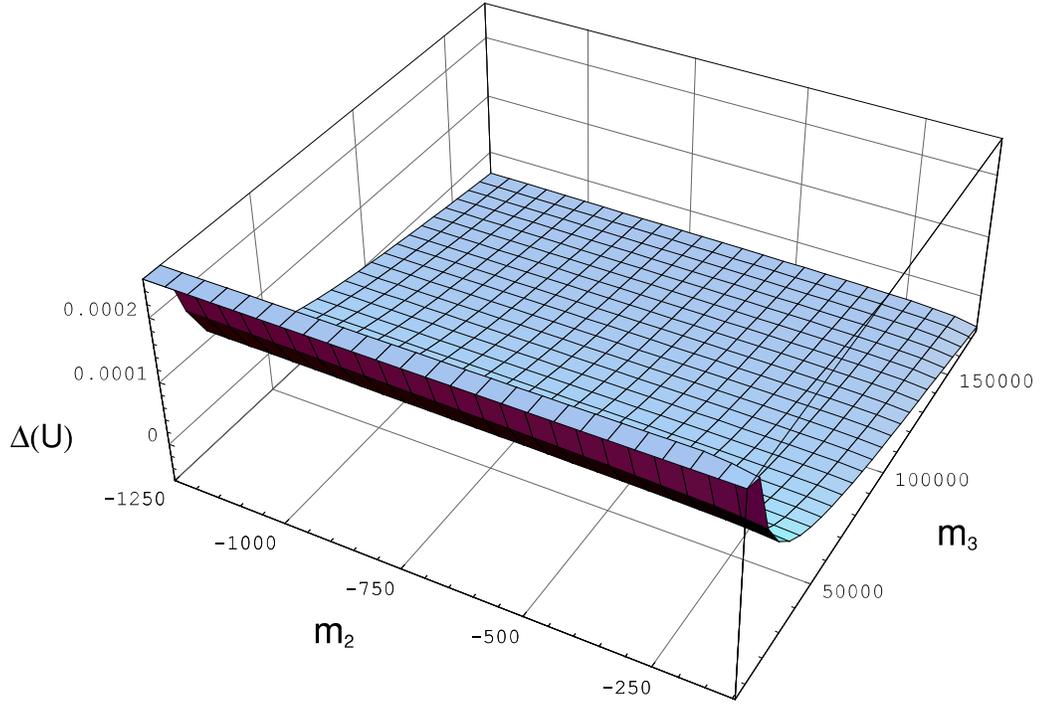}
\caption{\label{fig_2}Behaviour of $\Delta(U)$ as a function
of $m_2$ and $m_3$ for $m_1=3\,{\rm MeV}$. The intervals for
$m_2$ and $m_3$ are the same as in Fig.~1. $\Delta(U)$ increases
with the value of $m_1$. $\Delta(U)<0$ for the Case~A corner
(down--quark diagonal basis) where $|m_2|=m_c$ and $|m_3|=m_t$
and $\Delta(U)>0$ for the Case~B corner (up--quark diagonal
basis, $|m_2|=m_s$, $|m_3|=m_b$).} \end{figure}

\begin{figure}
\includegraphics*{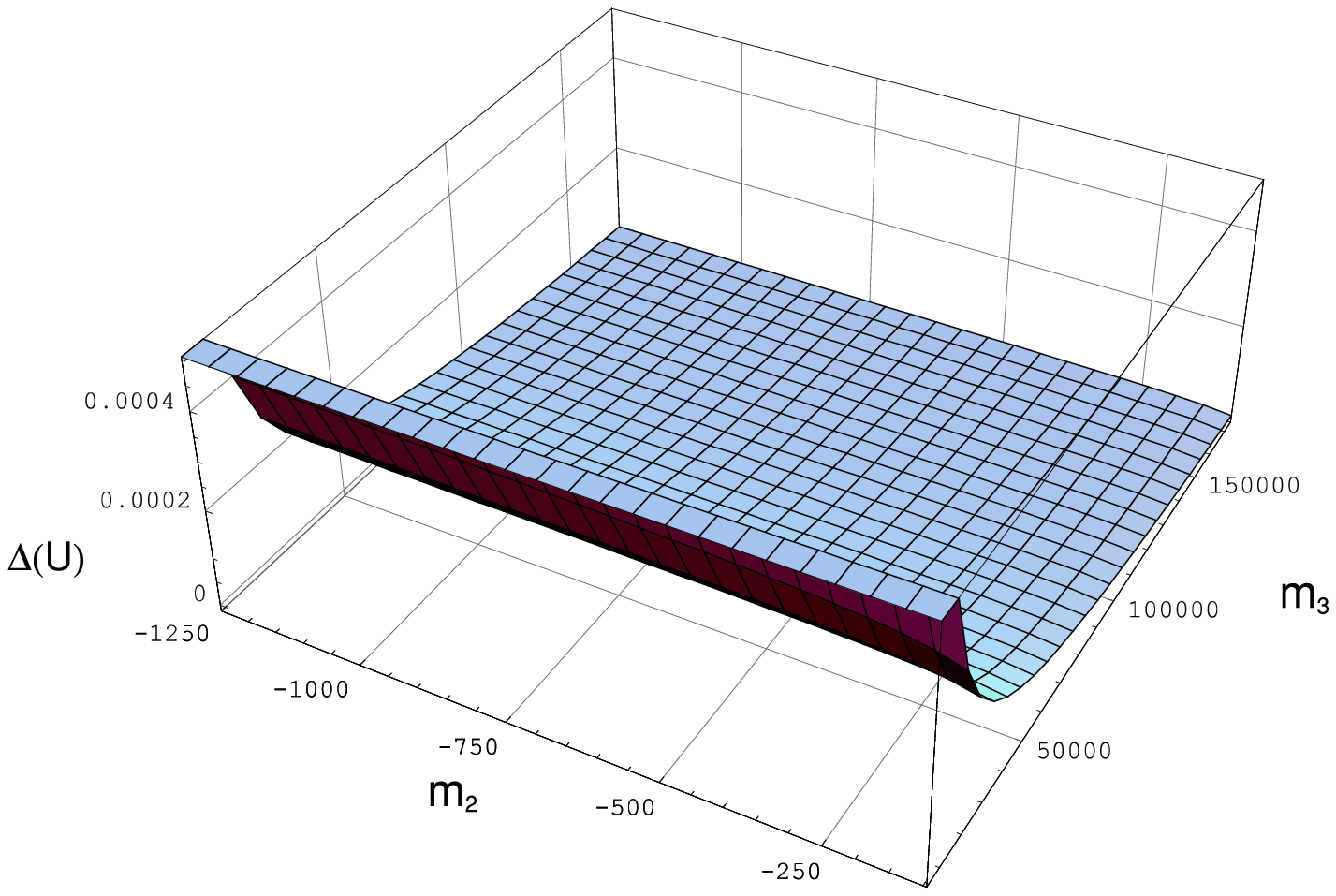}
\caption{\label{fig_3}Behaviour of $\Delta(U)$ as a function
of $m_2$ and $m_3$ for $m_1=7\,{\rm MeV}$. The intervals for
$m_2$ and $m_3$ are the same as in Fig.~1. $\Delta(U)$ increases
with the value of $m_1$ and now the whole graphic is positive.}
\end{figure}

We have also examined the dependence of $\Delta(U)$ as a function
of $m_2$ and $m_3$ for three typical values of $m_1$ according to
experimental $m_{\rm u}$ and $m_{\rm d}$~\cite{3}. The results are displayed
in Figures~\ref{fig_1}--\ref{fig_3}.

In general, we observe from Figs.~\ref{fig_1}--\ref{fig_3} that
the algebraic value of $\Delta(U)$ increases with the value of
$m_1$ in the selected range of values of $m_2$ and $m_3$, from
$-m_{\rm s}=-95\,{\rm MeV}$ to $-m_{\rm c}=-1.25\,{\rm GeV}$ and from
$m_{\rm b}=4.20\,{\rm GeV}$ to $m_{\rm t}=174.2\,{\rm GeV}$, respectively.
When $m_1=1.5\,{\rm MeV}$ (see Fig.~\ref{fig_1}) and also when
$m_1=3\,{\rm MeV}$ (see Fig.~\ref{fig_2}), $\Delta(U)<0$ for the
Case~A corner (down--quark diagonal basis) where
$|m_2|=m_{\rm c}$ and $|m_3|=m_{\rm t}$ and
$\Delta(U)>0$ for the Case~B corner (up--quark diagonal basis,
$|m_2|=m_{\rm s}$, $|m_3|=m_{\rm b}$). For
$m_1=7\,{\rm MeV}$, $\Delta(U)$ is positive for the whole graphic
(see Fig.~\ref{fig_3}).

The increase in the algebraic value of $\Delta(U)$ with
increasing $m_1$ (for given $|m_2|$ and $m_3$) observed in the
graphs can be understood algebraically. For the given $M$, the
condition $\Delta(U)>0$ can be expressed, in general, as

\begin{equation}
L<R
\label{LR}
\end{equation}

\noindent
where

\begin{equation}
L\equiv m_1\,|m_2|\,c - m_3\,|a|^2,
\label{L}
\end{equation}

\begin{equation}
R\equiv m_1\,|b|^2 - |m_2|\,|d|^2.
\label{R}
\end{equation}

\noindent
For the choice $a,b>0$ and $d=\mp i|d|$, Eq.~(\ref{16})
determines $a$, while Eq.~(\ref{15}) determines $b^2+|d|^2$ and
$b|d|$ is given by Eq.~(\ref{12}) in terms of $J(U)$, $a$, and
the masses. Thus, we can determine $b^2$ and $|d|^2$
individually. We assume $b>|d|$ as indicated by the numerical
fits in both the cases\footnote{The condition for real $b$ and
$|d|$ can be approximated as $J(U)<\sqrt{m_1|m_2|}/(2m_3)$ and
is satisfied in both cases.}. Since $m_3\gg |m_2|$ and $m_1$, an
approximate expression for $L\approx -2m_1|m_2|^2$. Given the
values of $m_i$, numerically $L_{\rm A}= -1.204\times
10^{7}\,({\rm MeV})^3$ and $L_{\rm B}= -8.055\times 10^{4}\,({\rm
MeV})^3$.

For $R$ we obtain

\begin{eqnarray*}
2R &=& (m_1-|m_2|)(b^2+|d|^2) + (m_1+|m_2|)(b^2-|d|^2)\\
&=& (m_1-|m_2|)(b^2+|d|^2) + (m_1+|m_2|)\sqrt{(b^2+|d|^2)^2 -
(2b|d|)^2}.
\end{eqnarray*}

\noindent
Given the numerical values of the $m_i$, in either case, we can
approximate this by expanding the square root to the first order
to obtain

\begin{equation}
R \approx m_1(b^2+|d|^2) - (m_1+|m_2|)\frac{b^2|d|^2}{b^2+|d|^2}.
\label{Rapprox}
\end{equation}

\noindent
For the given masses, $L$ can be neglected in comparison with the
first term of $R$ since $b^2+|d|^2\approx |m_2|m_3$.
Consequently, the condition Eq.~(\ref{LR}) is effectively $R>0$.
Since $b^2+|d|^2\approx |m_2|m_3$, this implies ($m_1\ne
0$)\footnote{Note that if $m_1$ was exactly zero from the start
it would imply $a=J(U)=0$ reducing Eq.~(\ref{LR}) to
$0<-|m_2||d|^2$\,!\ Also, for $d=0$ again $J(U)=0$ but
Eq.~(\ref{LR}) or Eq.~(\ref{condition}) is automatically
satisfied.}

\begin{equation}
\frac{m^2_1}{m^2_3}>J^2(U).
\label{condition}
\end{equation}

\noindent
The approximate algebraic condition Eq.~(\ref{condition}) gives
an insight into the numerical trend that $\Delta(U)$ increases
algebraically as $m_1$ increases.

\section{Concluding Remarks}

In this work we have examined constraints on mass matrices in the
quark sector  that arise due to measured properties of the mixing
matrix. Working in a basis where down--quark (up--quark) mass
matrix is diagonal and that the up--quark (down--quark) mass
matrix has a specific texture, we reconstruct the moduli of the
matrix elements of the mixing matrix taking the experimental
values of the quark masses and the Jarlskog invariant as inputs.
Comparing the modulii of the matrix elements of the mixing
matrix thus reconstructed with the available data, we find
better agreement for Case~B when the down--quark mass matrix has
the assumed form (see Eq.~(\ref{13})) with the up--quark mass
matrix diagonal rather than when the down-quark mass matrix
is diagonal (Case~A). This could well be attributed to the fact
that the mass ratios in the two cases are very different. It is
clear that in both cases one needs a more complicated mass matrix
than the $M$ considered above.

\begin{acknowledgments}

One of us (VG) is grateful to the School of Physics, University
of Hyderabad, India, for hospitality where a part of this work
was done. VG and G.~S-C would like to thank CONACyT (M\'exico)
for partial support.

\end{acknowledgments}

\begin{table*}

\caption{\label{table1}
Experimental and predicted numerical values of the asymmetry
$\Delta(V)$ (in units of $10^{-5}$). The calculated
$\Delta(V)$ is exactly the same for $|V_{12}|^2 -
|V_{21}|^2$, etc.\ (see remark after Eq.~(\ref{4})).
$\overline{\Delta(V)}$ in row~4 is the average of the three
values in rows~1 to 3. Case A: In down-quark diagonal basis,
with experimental values~\cite{3} of up-quark masses and $J(V)$
as inputs. Case B: In up-quark diagonal basis, with experimental
values~\cite{3} of down-quark masses and $J(V)$ as inputs.}

\begin{ruledtabular}

\begin{tabular}{c|c|c|c}

Quantity & Experiment\,\footnotemark[1] & Case
A\,\footnotemark[2] & Case B\,\footnotemark[2] \\

\hline

$|V_{12}|^2-|V_{21}|^2$ &
$5\pm 64$ & $6.2\pm 3.1$ & $110\pm 40$ \\

\hline

$|V_{23}|^2-|V_{32}|^2$ &
$5.1^{+1.3}_{-9.4}$ & $6.2\pm 3.1$ & $110\pm 40$ \\

\hline

$|V_{31}|^2-|V_{13}|^2$ &
$5.05^{+0.53}_{-1.04}$ & $6.2\pm 3.1$ & $110\pm 40$ \\

\hline

$\overline{\Delta(V)}$ & $5^{+21}_{-22}$ & $6.2\pm 3.1$ & $110\pm 40$ \\

\end{tabular}

\end{ruledtabular}

\footnotetext[1]{From Ref.~\cite{3}.}

\footnotetext[2]{From Eqs.~(\ref{7}) and (\ref{8}).}

\end{table*}

\begin{table}

\caption{\label{table2}
Experimental and predicted numerical values of the moduli of the
matrix elements $|V_{ij}|$ of the CKM-matrix $V$. Case A: In
down-quark diagonal basis, with experimental values~\cite{3} of
up-quark masses and $J(V)$ as inputs. Case B: In up-quark
diagonal basis, with experimental values~\cite{3} of down-quark
masses and $J(V)$ as inputs.}

\begin{ruledtabular}

\begin{tabular}{c|c|c|c}

Quantity & Experiment\,\footnotemark[1] & Case
A\,\footnotemark[2] & Case B\,\footnotemark[2] \\

\hline

$|V_{11}|$ & $0.97383^{+0.00024}_{-0.00023}$ & $0.9939\pm
0.0017$ & $0.975\pm 0.012$ \\

\hline

$|V_{12}|$ & $0.2272^{+0.0010}_{-0.0010}$ & $0.111\pm 0.015$ &
$0.224\pm0.051$ \\

\hline

$|V_{13}|$ & $0.00396^{+0.00009}_ {-0.00009}$ & $0.00360\pm
0.00060$ & $0.00123\pm 0.00013$ \\

\hline

$|V_{21}|$ & $0.2271^{+0.0010}_{-0.0010}$ & $0.110\pm 0.015$ &
$0.221\pm 0.051$ \\

\hline

$|V_{22}|$ & $0.97296^{+0.00024}_{-0.00024}$ & $0.9903\pm 0.0016$
& $0.964\pm 0.010$ \\

\hline

$|V_{23}|$ & $0.04221^{+0.00010}_{-0.00080}$ & $0.0843\pm 0.0031$
& $0.145\pm 0.020$ \\

\hline

$|V_{31}|$ & $0.00814^{+0.00032}_{-0.00064}$ & $0.0086\pm
0.0015$ & $0.0331\pm 0.0060$ \\

\hline

$|V_{32}|$ & $0.04161^{+0.00012}_{-0.00078}$ & $0.0839\pm
0.0031$ & $0.141\pm 0.020$ \\

\hline

$|V_{33}|$ & $0.999100^{+0.000034}_{-0.000004}$ & $0.99644\pm
0.00026$ & $0.9895\pm 0.0029$\\

\end{tabular}

\end{ruledtabular}

\footnotetext[1]{From Ref.~\cite{3}.}

\footnotetext[2]{From Eqs.~(\ref{5}) and~(\ref{4}).}

\end{table}

\begin{thebibliography}{}

\bibitem{2}
M.~Kobayashi and T.~Maskawa, Prog.\ Theor.\ Phys.\ D\ {\bf 35}
(1973) 652; N.~Cabibbo, Phys.\ Rev.\ Lett.\ {\bf 10} (1963) 531;
L.~L.~Chau and W.~-T.~Keung, Phys.\ Rev.\ Lett.\ {\bf 53} (1984)
1802.

\bibitem{1}
S.~Chaturvedi and V.~Gupta, Mod.\ Phys.\ Lett.\ A\ {\bf 21}
(2006) 907.

\bibitem{4}
C.~Jarlskog, Phys.\ Rev.\ Lett.\ {\bf 55} (1985) 1039.

\bibitem{3}
W.~-M.~Yao {\it et al.}, J.\ Phys.\ G:\ Nucl.\ Part.\ Phys.\
{\bf 33} (2006) 1.

\bibitem{7}
H.~Fritzsch, Phys.\ Lett.\ B\ {\bf70} (1977) 436,
Phys.\ Lett.\ B\ {\bf166} (1986) 423.

\bibitem{8}
T.~Kitazoe and K.~Tanaka, Phys.\ Rev.\ {\bf18} (1978) 3476;
H.~Georgi and D.~Nanapoulos, Phys.\ Lett.\ B\ {\bf82} (1979) 392;
B.~Stech, Phys.\ Lett.\ B\ {\bf 130} (1983) 189.

\end{thebibliography}
\end{document}